\documentclass{PoS}
\usepackage{amsmath,amssymb, bm}
\title{On the  physical DGLAP evolution of structure functions and its dependence on the renormalization scale\thanks{Work in collaboration with Marco Stratmann}}

\ShortTitle{Physical DGLAP evolution}

\author{\speaker{Martin Hentschinski}\\
Instituto de Ciencias Nucleares \\
Universidad Nacional Aut\'onoma de M\'exico\\
Apartado Postal 70-543 \\
M\'exico D.F. 04510 MX      
 \\
        E-mail: \email{hentschinski@correo.nucleares.unam.mx}}

      \abstract{Physical anomalous dimensions are a formulation of the DGLAP evolution of Deep Inelastic structure functions which is independent of factorization scheme and -scale. In this proceedings we provide an outlook on possible applications, in particular in the search of  saturation effects. As an original contribution we  present a short study of the renormalization scale dependence of physical evolved structure functions for large initial scale $Q_0^2=30$GeV$^2$}

\FullConference{XXIII International Workshop on Deep-Inelastic Scattering,\\
		27 April - May 1 2015\\
		Dallas, Texas}

\begin{document}

\section{Introduction}
\label{sec:intro}

Analysis of Deep-inelastic scattering (DIS) cross sections is
generally performed through global fits of scale-dependent quark and
gluon distribution functions $f_i(x,Q^2)$, $i=q,\bar{q},g$.  The
underlying theoretical framework of such analysis is based on the
collinear factorization theorem \cite{Collins:1989gx}, which organizes
the computation of DIS structure functions $F_{2,L}(x,Q^2)$ into the
convolution of short-distance Wilson coefficients and long-distance
parton distribution functions (PDFs) (see {\it e.g.}
\cite{Blumlein:2012bf} for a recent review). In order to formulate
such a theorem it is necessary to introduce a factorization scale
$\mu_f$ which separates long- and short-distance physics.
Independence of physical observables on $\mu_f$ allows then to derive
renormalization group equations (RGEs) which govern the scale
dependence of PDFs, known as the
Dokshitzer-Gribov-Lipatov-Altarelli-Parisi (DGLAP) equations.  Since
there is an infinite number of different ways to realize
factorization, one is left with an additional choice of the
factorization scheme with the $\overline{\mathrm{MS}}$ prescription
the generally adapted one. For observables, such as DIS structure functions,  any residual dependence on
factorization scheme and - scale $\mu_f$ is suppressed by an
additional power of $\alpha_s$, {\it i.e.}, is formally one order
higher in the perturbative expansion but not necessarily numerically
small.

\begin{figure}[ht]
  \centering
  \includegraphics[width=.9\textwidth]{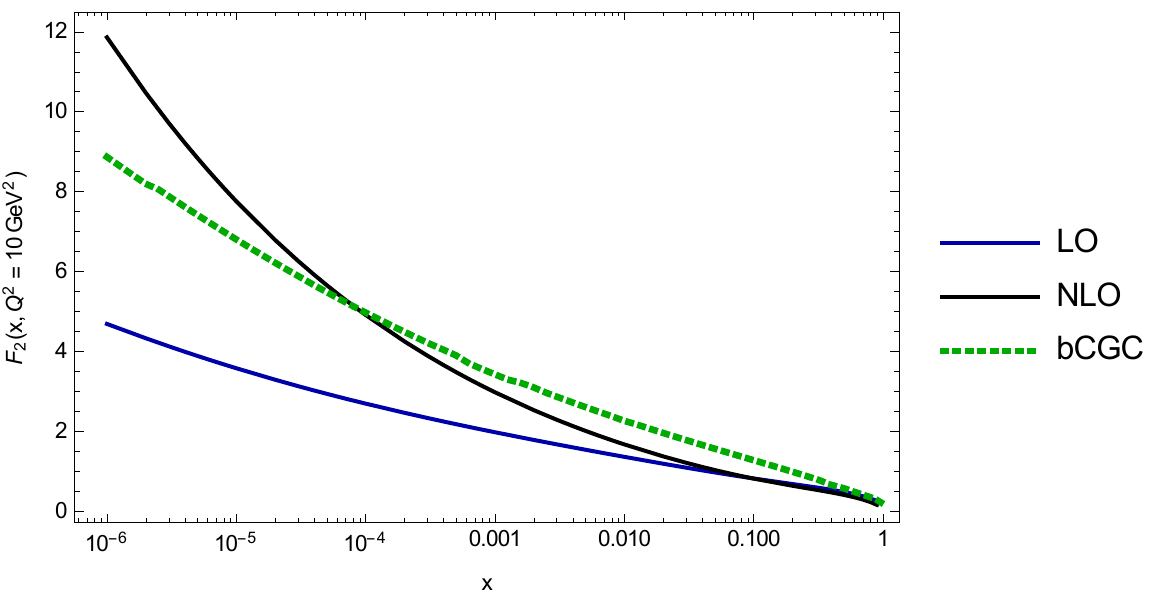}
  \includegraphics[width=.9\textwidth]{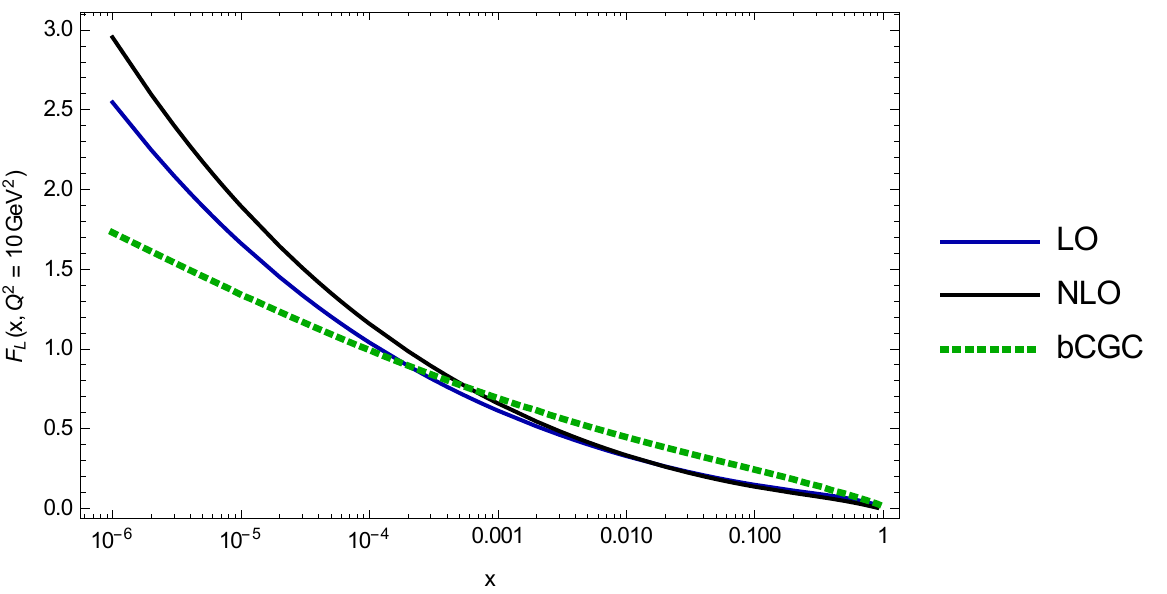}
  \caption{\it \small Gold structure functions $F_2$ and $F_L$ have
    been calculated at $Q^2=2$GeV$^2$ from the bCGC-model, using the
    fit \cite{Rezaeian:2013tka} to HERA proton DIS data, combined with  a scaling $Q_s^2 \to
    Q_S^2 A^{1/3}$. The result has been fitted and used as input for
    physical DGLAP evolution. The plots show a comparison of physical
    DGLAP evolution at leading (LO) and next-to-leading (NLO) order for the  doublet $(F_2, F_L)$ from  $Q_0^2=2$GeV$^2 \to Q^2=10$GeV$^2$ and the corresponding
    bCGC result at $Q^2=10$GeV$^2$.}
  \label{fig:bCGC}
\end{figure}

As an alternative to this conventional treatment it is also possible to formulate
QCD scale evolution equations directly for  observables
without referring to auxiliary, convention-dependent PDFs, which
avoids introduction of an artificial factorization scheme and -scale
dependence altogether \cite{Furmanski:1981cw}.  The
framework is suited best for theoretical analyses of DIS data; in
particular one remains in this case with the  renormalization scale
$\mu_r$ as the  only  theoretical ambiguity.  Since theory uncertainties are in this way reduced to a minimum,  it is therefore this  `physical' formulation of DGLAP evolution which is most suitable for extractions of  $\alpha_s$ from inclusive DIS data (for a first study see {\it e.g.} \cite{vanNeerven:1999ca}). Moreover, physical evolution allows for the most stringent tests of DGLAP evolution itself. This is of particular interest for regions of phase space where  a breakdown of collinear factorization is expected, such as the  limit $x \to 0$ of DIS structure functions,  where large parton densities eventually saturate and require a more complete description including terms  usually suppressed by powers of $Q^2$.  A potential application of physical evolution for such studies is illustrated and  discussed in Fig.~\ref{fig:bCGC}.
The outline of these proceedings is as follows: in
Sec.~\ref{sec:kernel} we provide some details on the derivation and
definition of physical evolution kernels while Sec.~\ref{sec:num}
presents an analysis of the remaining renormalization scale dependence
of physical evolution up to next-to-next-to-leading order (NNLO) in
$\alpha_s$ for large input scales $Q^2=30$GeV$^2$.  For more details
we refer the interested reader to \cite{preparation}.

\section{Physical evolution kernel}
\label{sec:kernel}

To define physical evolution kernels  we use that the  $x$-space  convolutions of coefficient functions and PDFs turn into  products in conjugate Mellin space,
%\begin{align}
 % \label{eq:Melli}
 $ a(n)  =  \int_0^1 d x x^{n-1} a(x) $.
%\end{align}
Moments of DIS structure functions $F_I(x, Q^2)$  can then be expressed as 
\begin{equation}
\label{eq:strfct}
F_{I}(n,Q^2) = 
\sum_k C_{I,k}\left(n,\alpha_s(\mu^2),\frac{Q^2}{\mu^2}, \frac{\mu_r^2}{\mu_f^2}\right)\,\cdot \, f_k\left(n,\alpha_s(\mu^2),\frac{\mu_f^2}{Q_0^2}, \frac{\mu_r^2}{\mu_f^2}\right).
 \end{equation}
The sum runs over all contributing quark flavors and the gluon,
each represented by a PDF $f_k$. The non-perturbative PDFs $f_k(n,\mu^2)$ obey the DGLAP evolution equations
\begin{equation}
\label{eq:dglap}
\frac{d f_k(n,\mu^2)}{d\ln \mu^2} = \sum_l P_{kl}(n,\alpha_s(\mu^2),\frac{Q^2}{\mu^2}) f_l(n,\mu^2),
\end{equation} 
while  coefficient functions $C_{I,k}$  \cite{vanNeerven:1991nn,Moch:1999eb,Kazakov:1987jk}  and splitting kernels $ P_{kl}$ \cite{ref:lo,ref:nlo,ref:nnlo} can be calculated in perturbative QCD and exhibit the following
expansion in $\alpha_s$
\begin{align}
\label{eq:pexpansion}
P_{kl}  &=  \sum_{m = 0}  \left( \frac{\alpha_s}{4 \pi}\right)^{1 + m} \; P^{(m)}_{kl}(n) \;
,& C_{I,k} &=  \sum_{m = 0} \left(\frac{\alpha_s}{4 \pi}\right)^{m_0 + m} \; C^{(m)}_{I,k}(n) \;,
\end{align}
where $m_0$ depends on the first non-vanishing order in $\alpha_s$ in
the expansion for the observable under consideration, {\it e.g.}
$m_0=0$ for $F_2$ and $m_0=1$ for $F_L$.  The DGLAP evolution equation
are formulated as $n_f -1$ evolution equations for the different
non-singlet quark flavor combinations and a $2 \times 2$ matrix valued
evolution equation, which evolves the flavor singlet vector $(\Sigma,
g)$; $ g(n, \mu^2)$ the gluon distribution and $ \Sigma(n, \mu^2) =
\sum_f^{n_f} \left[ q_f(n, \mu^2) + \bar{ q}_f(n, \mu^2) \right] $ the
quark flavor singlet. In the following we concentrate ourselves on the
flavor singlet sector only; for the physical evolution of the
non-singlet sector see {\it e.g.}  \cite{vanNeerven:1999ca}.  Using
that any doublet of flavor singlet observables $F=(F_A, F_B)$ is
related to the flavor singlet vector $(\Sigma, g)$ through a
coefficient matrix $C$
\begin{align}
  \label{eq:C}
  C & = \left(
  \begin{array}[h]{cc}
    C_{Aq} & C_{Ag} \\ C_{Bq} & C_{Bg}
  \end{array}
\right) &
P  & = \left(
  \begin{array}[h]{cc}
    P_{qq} & P_{qg} \\ P_{gq} & P_{gg}
  \end{array}
\right) &,
\end{align}
and introducing further a corresponding $2 \times 2$ matrix $P$ for the matrix-valued kernel of the DGLAP evolution in the flavor singlet sector, one finds in a straight forward manner
\begin{align}
  \label{eq:physev}
  \frac{d {F} (n, Q^2)}{ d \ln Q^2} & = \left( 4 \pi \beta \frac{d {C}}{ d \alpha_s} {C}^{-1}  + {C}\cdot  P \cdot {C}^{-1}  \right)\cdot  {F}.
\end{align}
where we made in addition use of the RG equation of $\alpha_s$ governed by the QCD beta function 
\begin{equation}
\label{eq:alphas}
\frac{d \alpha_s(\mu)}{d\ln \mu^2} = 4 \pi \beta(a_s) = - \alpha_s \sum_m \left(\frac{\alpha_s}{4 \pi}\right)^{m+1} \beta_m \, .
\end{equation}
The resulting  physical evolution kernels 
\begin{align}
  \label{eq:KQ2}
  {K}\left(\alpha_s(\mu_r^2), \frac{Q^2}{\mu_r^2} \right) & \equiv \left(  \beta \frac{d {C}}{ d a_s} {C}^{-1}  + {C}\cdot  P \cdot {C}^{-1}  \right) 
=
 \frac{\alpha_s(\mu_r^2)}{4 \pi}  \sum_{m = 0}  \left( \frac{\alpha_s(\mu_r^2)}{4 \pi}\right)^{m} \; K^{(m)}\left(n,  \frac{Q^2}{\mu_r^2} \right)  \; ,
\end{align}
are independent of factorization scheme and -scale
\cite{Furmanski:1981cw}  with the renormalization scale  $\mu_r$ as  their only remaining scale ambiguity at finite perturbative order.

\section{Renormalization scale dependence}
\label{sec:num}

\begin{figure}[ht]
  \centering
  \includegraphics[width=.45\textwidth]{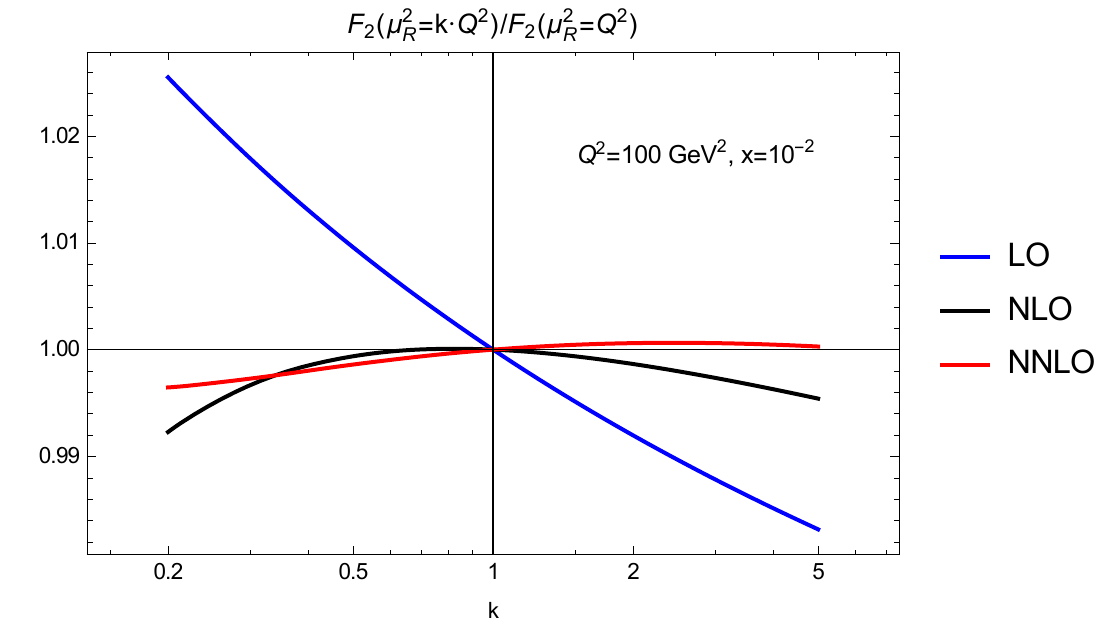}
 \includegraphics[width=.45\textwidth]{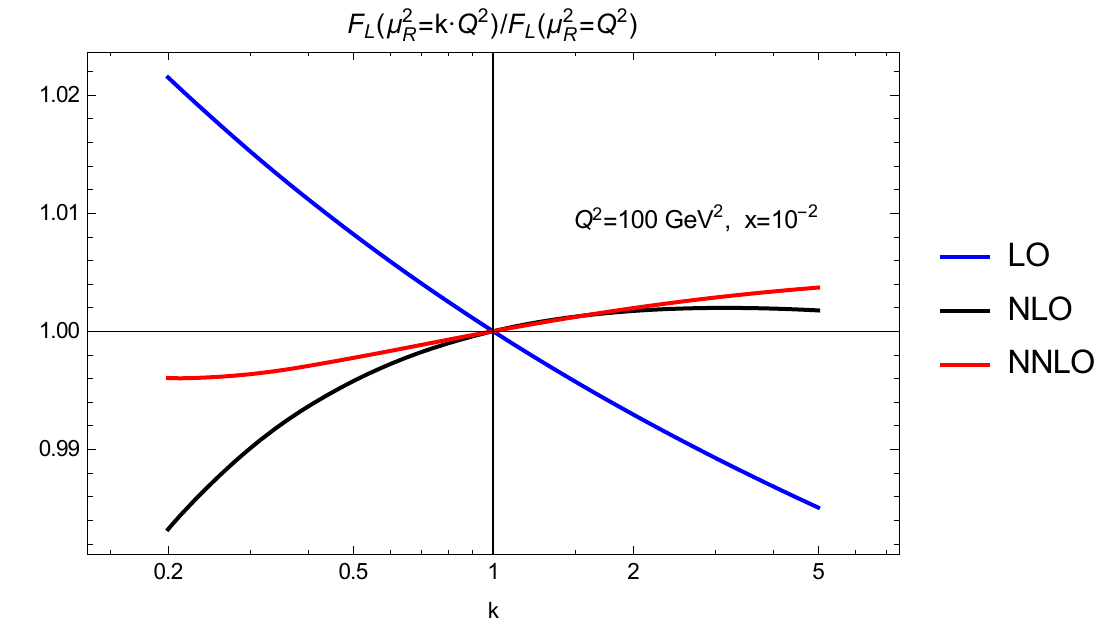} \\
 \includegraphics[width=.45\textwidth]{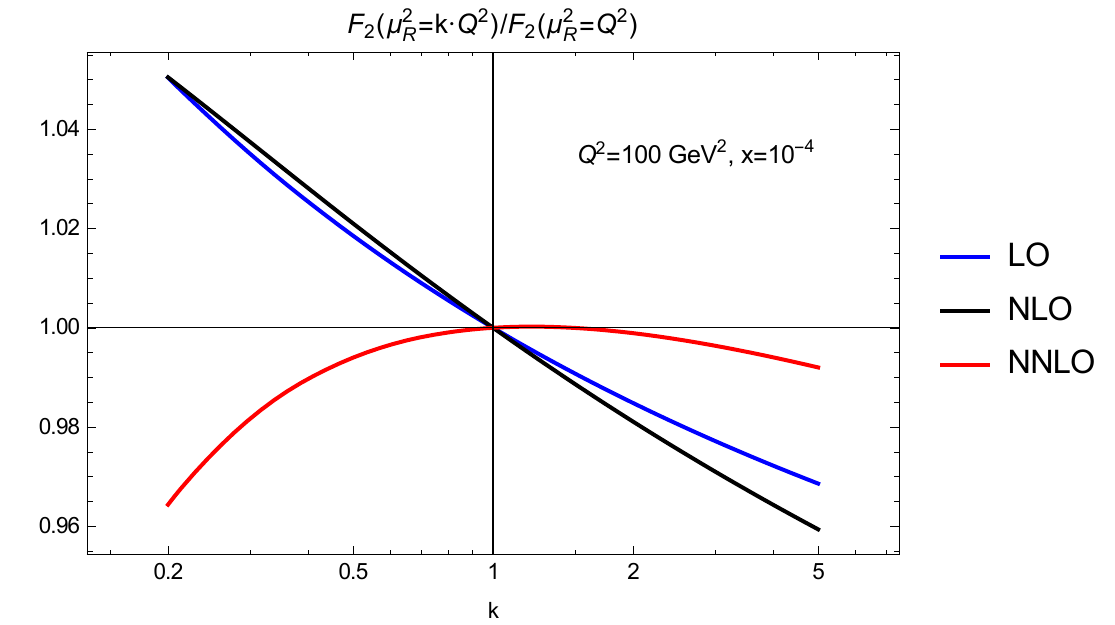}
 \includegraphics[width=.45\textwidth]{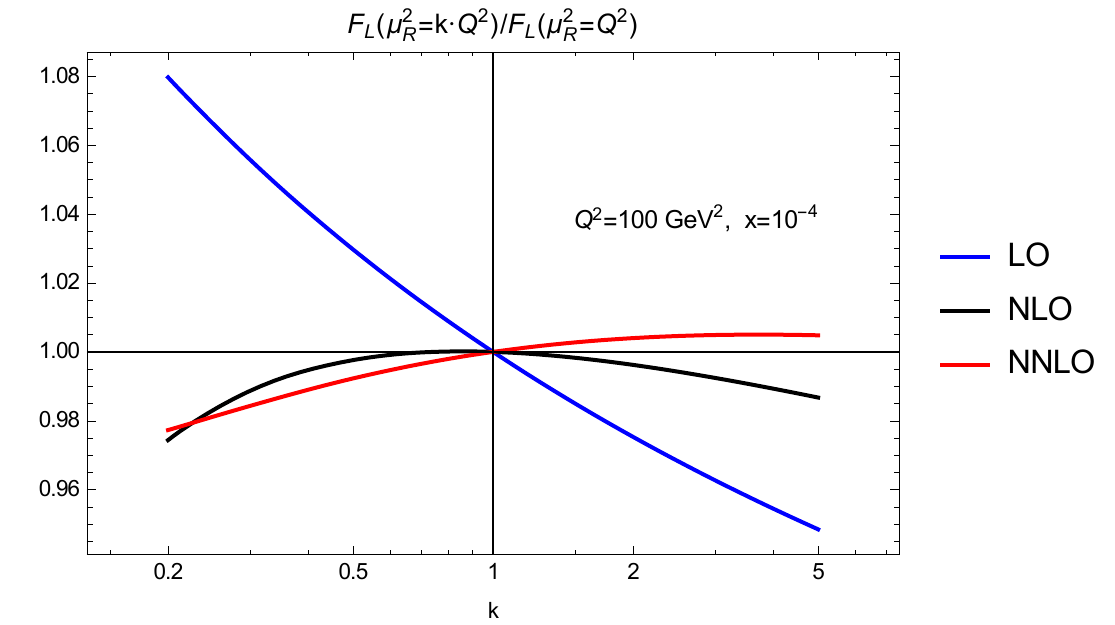} 
  \caption{\it Renormalization scale dependence of the doublet $(F_2, F_L)$}
  \label{fig:renom1}
\end{figure}
\begin{figure}[ht]
  \centering
  \includegraphics[width=.45\textwidth]{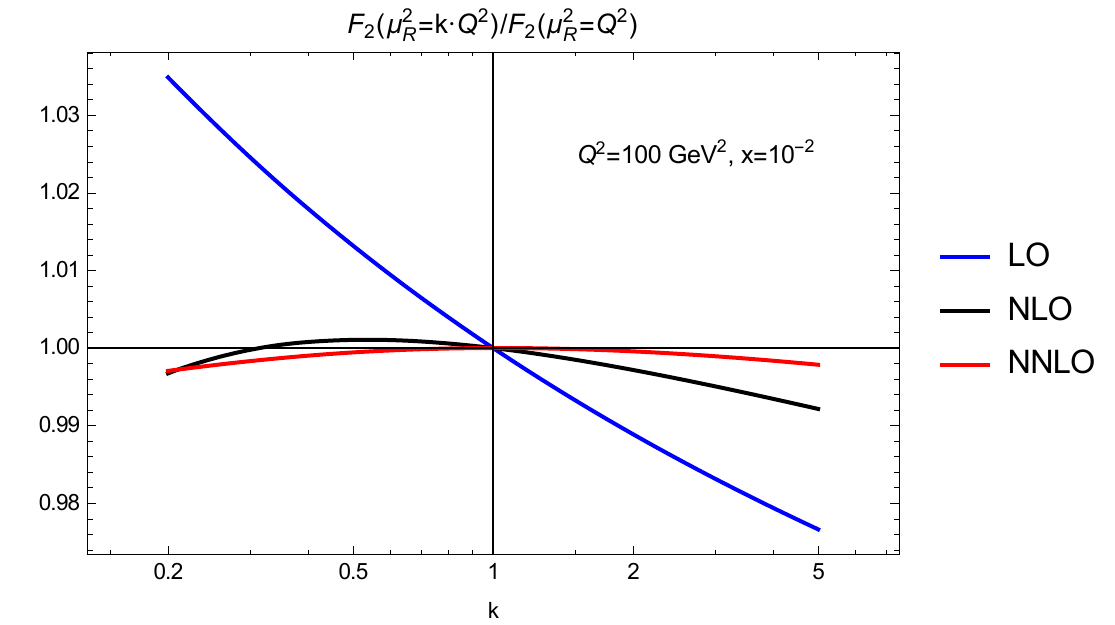}
 \includegraphics[width=.45\textwidth]{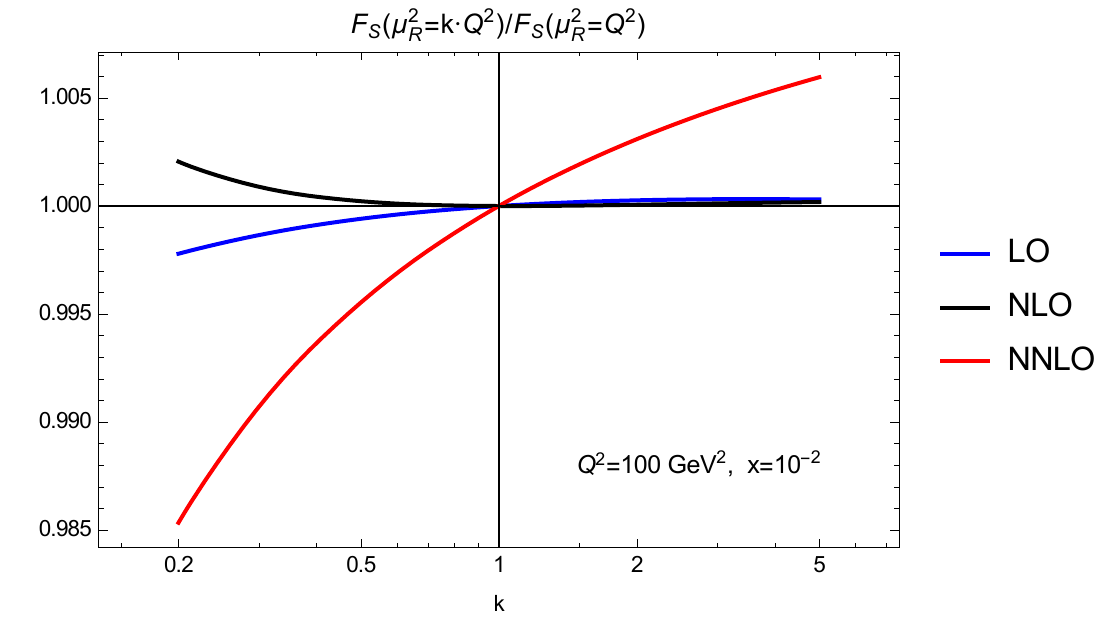} \\
 \includegraphics[width=.45\textwidth]{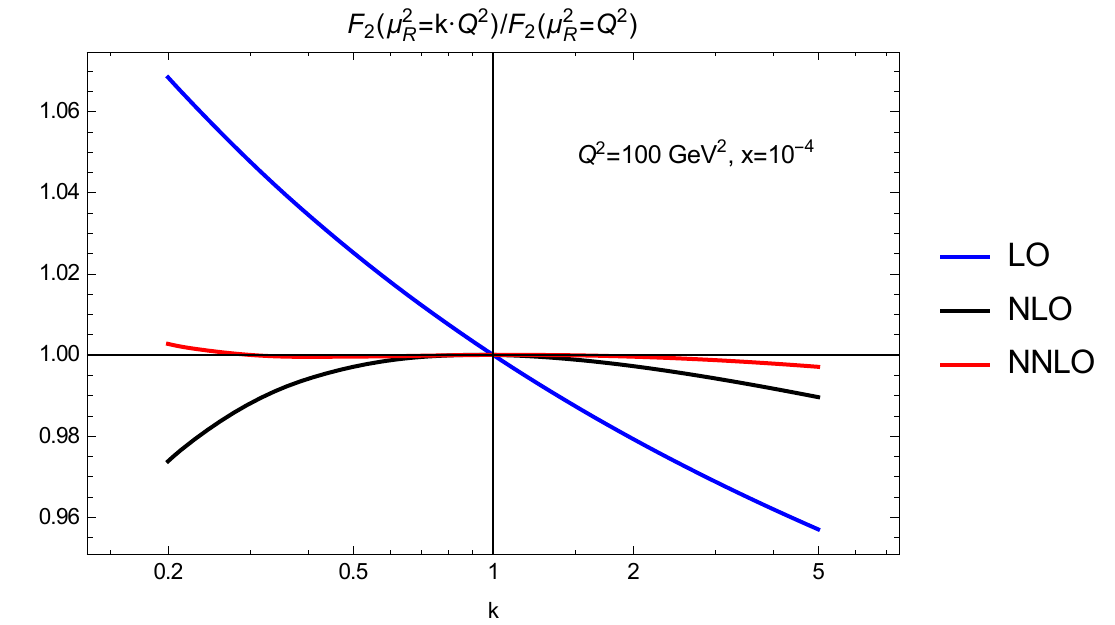}
 \includegraphics[width=.45\textwidth]{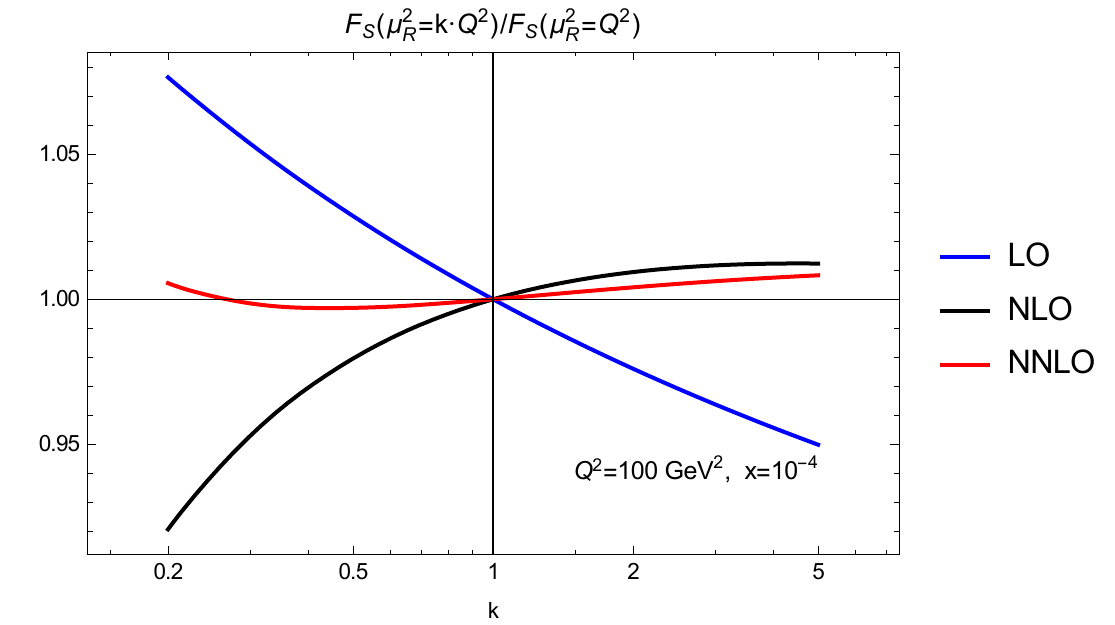} 
  \caption{\it Renormalization scale dependence  of the doublet $(F_2, F_S)$}
  \label{fig:renom2}
\end{figure}
To determine the renormalization scale dependence of physical evolution kernels we use kernels calculated at $\mu_r=Q$, see {\it e.g.} \cite{preparation}, and recover their full renormalization scale dependence using the same prescription as for conventional DGLAP splitting kernels {\it i.e.} by Taylor expanding  $\alpha_s(Q^2)$ in terms of $\alpha_s(\mu_r^2)$; see {\it e.g.} \cite{Vogt:2004ns} for a discussion in the case of splitting kernels. For the following numerical study we fix $n_f=4$  and use a realistic toy input at $Q^2=30$GeV$^2$ for quark singlet and gluon distribution (taken from \cite{Moch:2004xu}), 
\begin{align}
  \label{eq:1}
  x \Sigma(x) &= 0.6\, x^{-0.3}(1-x)^{3.5}(1+5\,x^{0.8})  \notag \\
  x g(x) &= 1.6\, x^{-0.3}(1-x)^{4.5}(1-0.6\, x^{0.3}) \; ,
\end{align}
 from which  we  calculate structure functions using LO coefficients, independent of the actual studied perturbative order. The strong coupling is obtained from solving Eq.~\eqref{eq:alphas} with the QCD beta function truncated at the corresponding perturbative order and  initial conditions fixed to  $\alpha_s(Q_0)=0.2$. We study as examples the flavor singlet sector of the doublets $(F_2, F_L)$ and $(F_2, F_S)$ with $F_S$ the $F_2$ scaling violations
\begin{align}
  \label{eq:2}
  F_S (x, Q^2) \equiv \frac{d F_2(x, Q^2)}{\ln Q^2}\; .
\end{align}
Our result are depicted in Fig.~\ref{fig:renom1} and Fig.~\ref{fig:renom2} and reveal a very small dependence on the chosen renormalization scale at NNLO, with the doublet  $(F_2, F_S)$ slightly less sensitive to the  variation of the renormalization scale. Note that the variations have been performed  over a very wide range {\it i.e.} $\mu_{r,0}^2/\text{GeV}^2 \in [6, 150]$ and $\mu_r^2/\text{GeV}^2 \in [20, 500]$ respectively.\\
 In conclusion we find a very mild dependence on the renormalization scale, if the initial scale for  DGLAP evolution is  rather large. For the case of small initial scales we refer to the discussion in  \cite{preparation}.

%%%%%%%%%%%%%%%%
\section*{Acknowledgments}
%%%%%%%%%%%%%%%%
%
This work was supported in part by UNAM-DGAPA-PAPIIT grant number 101515
and CONACyT-Mexico grant number 128534.

%%%%%%%%%%%%%%%%%%%%%%%%%%%


\begin{thebibliography}{99}
%%%%%%%%%%%%%%%%%%%%%%%%%%%
%\cite{Collins:1989gx}
\bibitem{Collins:1989gx} 
  J.~C.~Collins, D.~E.~Soper and G.~F.~Sterman,
  %``Factorization of Hard Processes in QCD,''
  Adv.\ Ser.\ Direct.\ High Energy Phys.\  {\bf 5}, 1 (1989)
  [hep-ph/0409313].
  %%CITATION = HEP-PH/0409313;%%
  %552 citations counted in INSPIRE as of 13 juil. 2015

%\cite{Blumlein:2012bf}
\bibitem{Blumlein:2012bf} 
  J.~Blumlein,
  %``The Theory of Deeply Inelastic Scattering,''
  Prog.\ Part.\ Nucl.\ Phys.\  {\bf 69}, 28 (2013)
  [arXiv:1208.6087 [hep-ph]].
  %%CITATION = ARXIV:1208.6087;%%
  %27 citations counted in INSPIRE as of 13 juil. 2015
%%%% phys. anom
\bibitem{Furmanski:1981cw} 
  W.~Furmanski and R.~Petronzio,
  %``Lepton - Hadron Processes Beyond Leading Order in Quantum Chromodynamics,''
  Z.\ Phys.\ C {\bf 11}, 293 (1982),
 	 %%CITATION = ZEPYA,C11,293;%%
%\bibitem{Catani:1996sc} 
  S.~Catani,
  %``Physical anomalous dimensions at small x,''
  Z.\ Phys.\ C {\bf 75}, 665 (1997),
    %%CITATION = HEP-PH/9609263;%%
%\bibitem{Blumlein:2000wh} 
  J.~Blumlein, V.~Ravindran, and W.~L.~van Neerven,
  %``On the Drell-Levy-Yan relation to O(alpha(s)**2),''
  Nucl.\ Phys.\ B {\bf 586}, 349 (2000).
	  %%CITATION = HEP-PH/0004172;%%
%
%\cite{vanNeerven:1999ca}
\bibitem{vanNeerven:1999ca} 
  W.~L.~van Neerven and A.~Vogt,
  %``NNLO evolution of deep inelastic structure functions: The Nonsinglet case,''
  Nucl.\ Phys.\ B {\bf 568}, 263 (2000)
  [hep-ph/9907472].
  %%CITATION = HEP-PH/9907472;%%
  %153 citations counted in INSPIRE as of 13 juil. 2015

 \bibitem{preparation}
M.~Hentschinski and M.~Stratmann,  %``On the Practical Application of Physical Anomalous Dimensions,''
  arXiv:1311.2825 [hep-ph].
  %%CITATION = ARXIV:1311.2825;%%


%\cite{Rezaeian:2013tka}
\bibitem{Rezaeian:2013tka}
  A.~H.~Rezaeian and I.~Schmidt,
  %``Impact-parameter dependent Color Glass Condensate dipole model and new combined HERA data,''
  Phys.\ Rev.\ D {\bf 88} (2013) 074016
  [arXiv:1307.0825 [hep-ph]].
  %%CITATION = ARXIV:1307.0825;%%
  %22 citations counted in INSPIRE as of 14 juil. 2015








\bibitem{vanNeerven:1991nn} 
  W.~L.~van Neerven and E.~B.~Zijlstra,
  %``Order alpha-s**2 contributions to the deep inelastic Wilson coefficient,''
  Phys.\ Lett.\ B {\bf 272}, 127 (1991);
	  %%CITATION = PHLTA,B272,127;%%
  %``Contribution of the second order gluonic Wilson coefficient to the deep inelastic structure function,''
  Phys.\ Lett.\ B {\bf 273}, 476 (1991);
  	%%CITATION = PHLTA,B273,476;%%
  %``Order alpha-s**2 QCD corrections to the deep inelastic proton structure functions F2 and F(L),''
  Nucl.\ Phys.\ B {\bf 383}, 525 (1992).
	  %%CITATION = NUPHA,B383,525;%%
\bibitem{Moch:1999eb} 
  S.~Moch and J.~A.~M.~Vermaseren,
  %``Deep inelastic structure functions at two loops,''
  Nucl.\ Phys.\ B {\bf 573}, 853 (2000).
	  %%CITATION = HEP-PH/9912355;%%
%
\bibitem{Kazakov:1987jk} 
  D.~I.~Kazakov and A.~V.~Kotikov,
  %``TOTAL ALPHA-s CORRECTION TO DEEP INELASTIC SCATTERING CROSS-SECTION RATIO, R = sigma-L / sigma-t IN QCD. CALCULATION OF LONGITUDINAL STRUCTURE FUNCTION,''
  Nucl.\ Phys.\ B {\bf 307}, 721 (1988)
  [Erratum-ibid.\ B {\bf 345}, 299 (1990)];
  	%%CITATION = NUPHA,B307,721;%%
  D.~I.~Kazakov, A.~V.~Kotikov, G.~Parente, O.~A.~Sampayo, and J.~Sanchez Guillen,
  %``Complete quartic (alpha(s)**2) correction to the deep inelastic longitudinal structure function F(L) in QCD,''
  Phys.\ Rev.\ Lett.\  {\bf 65}, 1535 (1990)
  [Erratum-ibid.\  {\bf 65}, 2921 (1990)];
	  %%CITATION = PRLTA,65,1535;%%
  J.~Sanchez Guillen, J.~Miramontes, M.~Miramontes, G.~Parente, and O.~A.~Sampayo,
  %``Next-to-leading order analysis of the deep inelastic R = sigma-L / sigma-total,''
  Nucl.\ Phys.\ B {\bf 353}, 337 (1991).
	  %%CITATION = NUPHA,B353,337;%%
%
%%%%%LO kernels
%
\bibitem{ref:lo} 
  D.~J.~Gross and F.~Wilczek,
  %``Asymptotically Free Gauge Theories. 2.,''
  Phys.\ Rev.\ D {\bf 9}, 980 (1974);
  	%%CITATION = PHRVA,D9,980;%%
%
  H.~Georgi and H.~D.~Politzer,
  %``Electroproduction scaling in an asymptotically free theory of strong interactions,''
  Phys.\ Rev.\ D {\bf 9}, 416 (1974);
  %%CITATION = PHRVA,D9,416;%%
%
  V.~N.~Gribov and L.~N.~Lipatov,
  %``Deep inelastic e p scattering in perturbation theory,''
  Sov.\ J.\ Nucl.\ Phys.\  {\bf 15}, 438 (1972)
  [Yad.\ Fiz.\  {\bf 15}, 781 (1972)];
	 %%CITATION = SJNCA,15,438;%%
%	 
  G.~Altarelli and G.~Parisi,
  %``Asymptotic Freedom in Parton Language,''
  Nucl.\ Phys.\ B {\bf 126}, 298 (1977);
 	 %%CITATION = NUPHA,B126,298;%%
  Y.~L.~Dokshitzer,
  %``Calculation of the Structure Functions for Deep Inelastic Scattering and e+ e- Annihilation by Perturbation Theory in Quantum Chromodynamics.,''
  Sov.\ Phys.\ JETP {\bf 46}, 641 (1977)
  [Zh.\ Eksp.\ Teor.\ Fiz.\  {\bf 73}, 1216 (1977)];
	  %%CITATION = SPHJA,46,641;%%
%
  K.~J.~Kim and K.~Schilcher,
  %``Scaling Violation in the Infinite Momentum Frame,''
  Phys.\ Rev.\ D {\bf 17}, 2800 (1978).
  	%%CITATION = PHRVA,D17,2800;%%

%
%%%%%%%NLO kernels
%
\bibitem{ref:nlo} 
  E.~G.~Floratos, D.~A.~Ross, and C.~T.~Sachrajda,
  %``Higher Order Effects in Asymptotically Free Gauge Theories: The Anomalous Dimensions of Wilson Operators,''
  Nucl.\ Phys.\ B {\bf 129}, 66 (1977)
  [Erratum-ibid.\ B {\bf 139}, 545 (1978)];
  	%%CITATION = NUPHA,B129,66;%%
%
  G.~Curci, W.~Furmanski, and R.~Petronzio,
  %``Evolution of Parton Densities Beyond Leading Order: The Nonsinglet Case,''
  Nucl.\ Phys.\ B {\bf 175}, 27 (1980);
	  %%CITATION = NUPHA,B175,27;%%
%
  W.~Furmanski and R.~Petronzio,
  %``Singlet Parton Densities Beyond Leading Order,''
  Phys.\ Lett.\ B {\bf 97}, 437 (1980);
	  %%CITATION = PHLTA,B97,437;%%  
%  
  A.~Gonzalez-Arroyo, C.~Lopez, and F.~J.~Yndurain,
  %``Second Order Contributions to the Structure Functions in Deep Inelastic Scattering. 1. Theoretical Calculations,''
  Nucl.\ Phys.\ B {\bf 153}, 161 (1979);
	  %%CITATION = NUPHA,B153,161;%%
%
  A.~Gonzalez-Arroyo and C.~Lopez,
  %``Second Order Contributions to the Structure Functions in Deep Inelastic Scattering. 3. The Singlet Case,''
  Nucl.\ Phys.\ B {\bf 166}, 429 (1980);
	  %%CITATION = NUPHA,B166,429;%%
%
  E.~G.~Floratos, R.~Lacaze, and C.~Kounnas,
  %``Space and Timelike Cut Vertices in QCD Beyond the Leading Order. 1. Nonsinglet Sector,''
  Phys.\ Lett.\ B {\bf 98}, 89 (1981);
	  %%CITATION = PHLTA,B98,89;%%
  %``Space And Timelike Cut Vertices In Qcd Beyond The Leading Order. 2. The Singlet Sector,''
  Phys.\ Lett.\ B {\bf 98}, 285 (1981);
    %``Higher Order QCD Effects in Inclusive Annihilation and Deep Inelastic Scattering,''
  Nucl.\ Phys.\ B {\bf 192}, 417 (1981).
%
%%%%NNLO kernels
%
\bibitem{ref:nnlo} 
  S.~Moch, J.~A.~M.~Vermaseren, and A.~Vogt,
  %``The Three loop splitting functions in QCD: The Nonsinglet case,''
  Nucl.\ Phys.\ B {\bf 688}, 101 (2004);
  	%%CITATION = HEP-PH/0403192;%%
  %``The Three-loop splitting functions in QCD: The Singlet case,''
  Nucl.\ Phys.\ B {\bf 691}, 129 (2004).
	  %%CITATION = HEP-PH/0404111;%%  
%



\bibitem{Vogt:2004ns} 
  A.~Vogt,
  %``Efficient evolution of unpolarized and polarized parton distributions with QCD-PEGASUS,''
  Comput.\ Phys.\ Commun.\  {\bf 170}, 65 (2005).
	  %%CITATION = HEP-PH/0408244;%%


%\cite{Moch:2004xu}
\bibitem{Moch:2004xu} 
  S.~Moch, J.~A.~M.~Vermaseren and A.~Vogt,
  %``The Longitudinal structure function at the third order,''
  Phys.\ Lett.\ B {\bf 606}, 123 (2005)
  [hep-ph/0411112].
  %%CITATION = HEP-PH/0411112;%%
  %123 citations counted in INSPIRE as of 13 juil. 2015



\end{thebibliography}
\end{document}